\documentclass[twocolumn,
aps,nofootinbib,showpacs,showkeys,preprint
tightenlines,preprintnumbers,]{revtex4}
\usepackage{amsfonts}
\usepackage{epsf,epsfig,subfigure,graphicx,amsmath,amssymb}
\usepackage{color}

\def\E6{{\rm E_6}}
\def\EE8{{\rm E_8\times E_8}}

\input{tcilatex}
\begin{document}

\title{\textbf{On Discrete Gauge Symmetries in Trinification Model} }
\author{R. Ahl Laamara$^{1,2}$\thanks{$\ast $ ahllaamara@gmail.com }, M.
Drissi El Bouzaidi$^{1}$\thanks{
$\dag $ mohcinebouzaidi@gmail.com }, S.-E. Ennadifi$^{1}$\thanks{
$\ddag $ ennadifis@gmail.com }, S. Nassiri$^{1}$\thanks{
$\S $ nass.sami@gmail.com }}
\affiliation{{\small $^{1}$LHEP-MS, Faculty of Science, University Mohammed V, Rabat,
Morocco}\\
{\small $^{2}$Center of Physics and Mathematics, Rabat, Morocco}}

\begin{abstract}
Given the important role of discrete gauge symmetries in viable models, we
discuss these symmetries in intersecting D6-brane trinification model where
the $Z_{N}$ symmetry\ is investigated and its identification is shown.
\end{abstract}

\email{nass.sami@gmail.com}
\maketitle

\section{Introduction}

Recently, intersecting D-brane models in string theory have attracted much
attention in viable particle physics models \cite{1,2,3,4,5}. Considerable
works in this direction have been devoted to Standard-like Models (SM) \cite%
{6,7,8}. In these constructions, discrete gauge symmetrie seem to play
interesting role in realistic model buildings \cite{9,10}. Indeed, their
investigations have important implications, as they guarantee proton
stability, in agreement with experimental bounds \cite{11,12}.

Discrete symmetries are strongly constrained by anomaly cancellation
conditions \cite{9,10}. In particular abelian $Z_{N}$ discrete gauge
symmetries were imposed to forbid dangerous Lepton and Baryon number
violating operators, and can be realized as discrete remnants of continuous $%
U(1)$ gauge symmetries, with Ramon-Ramon (RR) 2-forms $B_{2}$ couplings to
the $U(1)$ gauge group field strengths $F_{a}$, ($BF$ couplings), that are
broken by scalars with charge $N$ under the respective $U(1)$ acquiring
vacuum expectation values (vev's) \cite{12,13,14}. One of the well studied
models is trinification $U(3)^{3}$ which can arise in the D-brane
constructions, where all fermions representations are charged under three
additional anomalous $U(1)^{\prime }s$ \cite{15,16,17,18}. One linear
combination of these $U(1)$ symmetries is anomaly free and can serve as a
hypercharge component, leading to very interesting phenomenological
implications \cite{15,16}.

In the present paper we discuss the origin of discrete gauge symmetries $%
Z_{N}$ in trinification model $U(3)_{C}\times U(3)_{L}\times U(3)_{R}$. We
start with a short description of the basic features of intersecting
D6-brane in type IIA constructions trinification model, and we briefly
review the basic considerations on the origin of a discrete $Z_{N}$
symmetries. Then we discuss the constraints and \ presence of these discrete
gauge symmetries\ imposed by the corresponding winding numbers of the model.

\section{Intersecting D6-branes and discrete symmetries}

\subsection{Intersecting D6-brane trinification model}

In this section, we review the results of the works where the authors
describe the gauge symmetry $\mbox{U(3)}_{C}\times \mbox{U(3)}_{L}\times %
\mbox{U(3)}_{R}$ in the context of intersecting D-branes on Type IIA
orientifolds \cite{15,16,17}. In these works, where three-stacks of
D6-branes filling out the four-dimensional space-time and wrapping
three-cycles $\pi _{a}$ in the internal Calabi--Yau threefold and moving on
orientifolded geometry, each stack contains 3 parallel D6-branes almost
coincident, and gives rise to the following gauge symmetry,

\begin{equation}
\begin{tabular}{ccccc}
$\mbox{U(3)}_{a}$ & $\times $ & $\mbox{U(3)}_{b}$ & $\times $ & $\mbox{U(3)}%
_{c}.$%
\end{tabular}
\label{eq1}
\end{equation}

The D-brane analogue of the trinification model construction involves
symmetries $\mbox{U(3)}=\mbox{SU(3)}\times \mbox{U(1)}$ and also contains
three extra $U(1)$ abelian symmetries, thus, the $\mbox{U(3)}^{3}$ gauge
group can be equivalently written as,

\begin{equation}
\begin{tabular}{ccccccccccc}
$\mbox{SU(3)}_{a}$ & $\times $ & $\mbox{SU(3)}_{b}$ & $\times $ & $%
\mbox{SU(3)}_{c}$ & $\times $ & $\mbox{U(1)}_{a}$ & $\times $ & $\mbox{U(1)}%
_{b}$ & $\times $ & $\mbox{U(1)}_{c}.$%
\end{tabular}
\label{eq2}
\end{equation}

The all three abelian $\mbox{U(1)}_{a,c,b}$ factors have mixed anomalies
with the non-abelian $\mbox{SU(3)}_{a}\times \mbox{SU(3)}_{b}\times %
\mbox{SU(3)}_{c}$ gauge parts which are determined by the contributions of
three fermion generations are proportional to $A\sim TrQ_{I}T_{J}^{2}$ where 
$T_{J}=\left\{ SU\left( 3\right) _{C},SU\left( 3\right) _{L},SU\left(
3\right) _{R}\right\} $ and $Q_{I}=\left\{ U\left( 1\right) _{C},U\left(
1\right) _{L},U\left( 1\right) _{R}\right\} $ with,

\begin{equation}
A=\left( 
\begin{array}{ccc}
0 & +1 & -1 \\ 
-1 & 0 & +1 \\ 
+1 & -1 & 0%
\end{array}%
\right) .  \label{eq3}
\end{equation}

There is only one anomaly-free $U(1)$ combination, namely,

\begin{equation}
\begin{tabular}{ccccccc}
$U(1)_{Z}$ & $=$ & $U(1)_{C}$ & $+$ & $U(1)_{L}$ & $+$ & $U(1)_{R}.$%
\end{tabular}
\label{eq4}
\end{equation}

The remaining two orthogonal combinations $U(1)_{C}-U(1)_{R}$ and $%
U(1)_{C}-2U(1)_{L}+U(1)_{R}$ have anomalies which are canceled by a
generalized Green--Schwarz mechanism (GSM). Generally, some direct
phenomenological consequences in low energy physics comes from the
cancellation of $U(1)$ anomalies by means of a GSM. In particular, some
abelian gauge fields get a mass term and, and consequently, the
corresponding $U(1)$ factors must be removed from the gauge group. Since the
mechanism by these gauge bosons acquire a mass does not involve a
non-vanishing vev for a scalar field, these massive $U(1)$ symmetries will
remain as global symmetries of the effective Lagrangian.

It possible to define a hypercharge component,

\begin{equation}
\begin{tabular}{ccccccc}
$Y_{Z}$ & $=$ & $Q_{C}$ & $+$ & $Q_{L}$ & $+$ & $Q_{R}.$%
\end{tabular}
\label{eq5}
\end{equation}

The spectrum of the trinification model in the intersecting of 3 stacks of
D6-branes context involves two kinds of representations; those that are
obtained from strings with both ends attached to two different branes have
the following quantum numbers for quarks $Q$ and leptons $L$ \cite{15,16,17},

\begin{equation*}
\begin{tabular}{cccccccc}
$Q$ & $=$ & $\left( 3,\overline{3},1\right) _{\left( 1,-1,0\right) }$ & $,$
& $\overline{Q}$ & $=$ & $\left( \overline{3},1,3\right) _{\left(
-1,0,1\right) }$ & $,$%
\end{tabular}%
\end{equation*}

\begin{equation}
\begin{tabular}{lll}
$L$ & $=$ & $\left( 1,3,\overline{3}\right) _{\left( 0,1,-1\right) }$%
\end{tabular}%
,  \label{eq6}
\end{equation}%
and those whose both ends are on the same brane stack,

\begin{equation}
\begin{tabular}{ccccccc}
$H_{u}$ & $+$ & $H_{d}$ & $=$ & $\left( 1,3,\overline{3}\right) _{\left(
0,1,-1\right) }$ & $+$ & $\left( 1,\overline{3},3\right) _{\left(
0,-1,1\right) }.$%
\end{tabular}
\label{eq7}
\end{equation}

These Higgs states are sufficient to break the $\mbox{U(3)}^{3}\times %
\mbox{U(1)}^{3}$gauge symmetry. The multiple families arise from the
intersection of two D6-branes in multiple points on the internal space,
where the number of families is the topological intersection number of two
3-cycles in middle dimensional cohomology.

In the present setup, considering compactifications on a 6-torus factorized
as $T^{6}=T^{2}\times T^{2}\times T^{2}$, there are three-stacks of
D6-branes wrapped on 3-cycles, denoted $D6_{C}$ (color), $D6_{L}$ (left) and 
$D6_{R}$ (right)$.$ The left-handed quarks $\left( 3,\overline{3},1\right) $
are localized at the intersection of brane $D6_{C}$ and $D6_{L},$ while the
right handed quarks $\left( \overline{3},1,3\right) $ are localized at the
intersection of brane $D6_{C}$ and $D6_{R}$. The leptons $L=\left( 1,3,%
\overline{3}\right) $ arise at intersections of brane $D6_{L}$ and $D6_{R}.$

In the intersecting D-brane scenario, the number of intersections for a $%
D6_{a}-D6_{b}$ sector is given by the product of the intersections in each
of them,

\begin{equation*}
\begin{tabular}{cc}
$I_{ab}$ & $=$%
\end{tabular}%
\end{equation*}

\begin{equation}
\begin{tabular}{l}
$\left( m_{a1}n_{b1}-m_{b1}n_{a1}\right) \left(
m_{a2}n_{b2}-m_{b2}n_{a2}\right) \left( m_{a3}n_{b3}-m_{b3}n_{a3}\right) ,$%
\end{tabular}
\label{eq8}
\end{equation}

where $(n_{ai},m_{ai})$ the wrapping numbers of the $D6_{a}$ stack around
the the i$^{th}$ torus.

In addition the additional matter fields arise from the sectors $%
D6_{a}-\Omega RD6_{b}$, whose multiplicity are given by,

\begin{equation*}
\begin{tabular}{cc}
$I_{ab^{\ast }}$ & $=$%
\end{tabular}%
\end{equation*}

\begin{equation}
\begin{tabular}{l}
$\left( m_{a1}n_{b1}+m_{b1}n_{a1}\right) \left(
m_{a2}n_{b2}+m_{b2}n_{a2}\right) \left( m_{a3}n_{b3}+m_{b3}n_{a3}\right) ,$%
\end{tabular}
\label{eq9}
\end{equation}

and there are three antisymmetric representations $(3,1,1),(1,3,1),(1,1,3)$,

\begin{equation}
\begin{tabular}{lll}
$I_{aa^{\ast }}$ & $=$ & $m_{a1}m_{a2}m_{a3}.$%
\end{tabular}
\label{eq10}
\end{equation}

In our model building, their numbers are,

\begin{equation}
\begin{tabular}{ccccccccccc}
$I_{CL}$ & $=$ & $3,$ & $I_{CR}$ & $=$ & $-$ & $3$ & and & $I_{LR}$ & $=$ & $%
3,$%
\end{tabular}
\label{eq11}
\end{equation}%
which lead to further constraints on the winding numbers.

These are represented in the following table,

\begin{table}[th]
\begin{center}
\begin{equation*}
\begin{tabular}{||c||c||c||c||c||c||}
\hline\hline
Intersections & Matter fields & Representations & $Q_{C}$ & $Q_{L}$ & $Q_{R}$
\\ \hline\hline
$\left( CL\right) $ & $Q_{L}$ & $\left( 3,\overline{3},1\right) $ & $1$ & $%
-1 $ & $0$ \\ \hline\hline
$\left( CR\right) $ & $Q_{R}$ & $\left( \overline{3},1,3\right) $ & $-1$ & $%
0 $ & $1$ \\ \hline\hline
$\left( LR\right) $ & $L$ & $\left( 1,3,\overline{3}\right) $ & $0$ & $1$ & $%
-1$ \\ \hline\hline
$\left( LR\right) $ & $H$ & $\left( 1,3,\overline{3}\right) $ & $0$ & $1$ & $%
-1$ \\ \hline\hline
\end{tabular}%
\end{equation*}%
\end{center}
\caption{Fields content and their representations in Trinification model. }
\end{table}

In addition the RR tadpole cancellation conditions should also be taken into
account. The restrictions imposed on the $(n_{ai},m_{ai})$ sets originating
from these conditions read,

\begin{equation}
\begin{tabular}{ccccccccccc}
$T_{0}$ & $=$ & $\sum_{a=c,l,r}$ & $N_{a}n_{a1}n_{a2}n_{a3}$ & $=$ & $16$ & $%
\text{\ }i$ & $\neq $ & $j$ & $\neq $ & $k.$%
\end{tabular}
\label{eq12}
\end{equation}

It is very difficult to satisfy the RR tadpole cancellation conditions,
because of large RR charges from three $U(3)$ groups. Additional stacks with 
$U(1)$ groups and filler branes are also used to satisfy the RR tadpole
cancellation conditions \cite{15,16,17,18}.

In intersecting D-brane models the cubic non-abelian anomalies are cancelled
automatically when the RR tadpole cancellation conditions are satisfied. The
mixed anomalies $SU\left( N\right) _{b}^{2}-U\left( 1\right) $ appears when
the gauge group structure of the D-brane constructions involves symmetries $%
U(N)=SU(N)\times U(1)$ and does not vanish even after imposing the tadpole
conditions presented above. These anomalies are cancelled by a generalized
GSM which involves some set of $BF$ couplings to a set of RR 2-forms. In
particular, for one stack of $N_{a}$ D-brane with wrapping numbers $%
(n_{ai},m_{ai})$, the four-dimensional couplings of the $B_{2}^{i}$ to the $%
U(1)$ field strength $F_{a}$ read,

\begin{equation}
\begin{tabular}{cc}
$N_{a}m_{a1}m_{a2}m_{a3}\int_{M_{4}}B_{2}^{0}\wedge F_{a},$ & $\text{\ }%
N_{a}n_{aj}n_{ak}m_{ai}\int_{M_{4}}B_{2}^{i}\wedge F_{a},$%
\end{tabular}
\label{eq13}
\end{equation}%
which are of the form $\sum\nolimits_{i}c_{a}^{i}B_{2}^{i}\wedge F_{a}$ and $%
c_{a}^{0}=N_{a}m_{a1}m_{a2}m_{a3}$, $c_{a}^{i}=N_{a}n_{aj}n_{ak}m_{ai}$
depend on the winding numbers.

\subsection{Discrete gauge symmetries}

Discrete symmetries are well motivated in supersymmetric field theories from
a phenomenological point of view, as they guarantee proton stability in
agreement with experimental bounds. These symmetries, that are less
constrained by anomalies than their continuous counter parts, arise
naturally in string theory, as it will be shown in the next section. $Z_{N}$
symmetries thus offer excellent\ ingredients\ for viable models \cite%
{10,11,12,13,14}.

Generically, in field theory, discrete $Z_{N}$ symmetry acts on the chiral
superfelds $\Psi _{j}$ by a global phase rotation,

\begin{equation}
\begin{tabular}{cccc}
$\Psi _{j}$ & $\rightarrow $ & $e^{i\alpha _{j}2\pi /N}$ & $\Psi _{j},$%
\end{tabular}
\label{eq14}
\end{equation}%
where the integer charge $\alpha _{j}$ is defined modulo an integer shift $N$%
,

\begin{equation}
\begin{tabular}{ccccccccc}
$\alpha _{j}$ & $\sim $ & $\alpha _{j}$ & $+$ & $m_{j}N$ & $\alpha _{j},$ & $%
m_{j}$ & $\in $ & $%
%TCIMACRO{\U{2124} }%
%BeginExpansion
\mathbb{Z}
%EndExpansion
.$%
\end{tabular}
\label{eq15}
\end{equation}

More specifically, discrete symmetries offer excellent extensions to the
Supersymmetric SM in order to prohibit the lepton-and/or baryon-number
violating operators. The possible $Z_{N}$ generation independent discrete
symmetries of the Supersymmetric SM with generator $g_{N}$ can be expressed
in terms of products of powers of three mutually discrete generators $%
A_{N}=e^{i2\pi A/N},L_{N}=e^{i2\pi L/N}$ and $R_{N}=e^{i2\pi R/N}$,

\begin{equation}
\begin{tabular}{ccccccc}
$g_{N}$ & $=$ & $A_{N}^{n}$ & $\times $ & $L_{N}^{p}$ & $\times $ & $%
R_{N}^{m},$%
\end{tabular}
\label{eq16}
\end{equation}%
where the exponents run over $m,n,p=0,1,...N-1.$ The most general $Z_{N}$
symmetry allowing for the presence of the charges of the chiral
Supersymmetric SM matter fields are given in table 2,

\begin{center}
\begin{table}[th]
\begin{center}
\begin{equation*}
\begin{tabular}{||c||c||c||c||c||c||c||c||c||}
\hline\hline
$Generator$ & $Q_{L}$ & $\overline{U}_{R}$ & $\overline{D}_{R}$ & $L$ & $%
\overline{E}_{R}$ & $\overline{N}_{R}$ & $H_{u}$ & $H_{d}$ \\ \hline\hline
$R$ & $0$ & $n-1$ & $1$ & $0$ & $1$ & $n-1$ & $1$ & $n-1$ \\ \hline\hline
$L$ & $0$ & $0$ & $0$ & $n-1$ & $1$ & $1$ & $0$ & $0$ \\ \hline\hline
$A$ & $0$ & $0$ & $n-1$ & $n-1$ & $0$ & $1$ & $0$ & $1$ \\ \hline\hline
\end{tabular}%
\end{equation*}%
\end{center}
\caption{Chiral fields and their discrete $R$, $L$, $A$ charges assignment. }
\end{table}
\end{center}

Given this assignment the Supersymmetric SM particles carry discrete charges
under a $g_{N}$ transformation.

\begin{center}
\begin{table}[th]
\begin{center}
$%
\begin{tabular}{||c||c||c||c||c||c||c||c||c||}
\hline\hline
$q_{Q_{L}}$ & $=$ & $0$ & $q_{U_{R}}$ & $=$ & $-m$ & $q_{D_{R}}$ & $=$ & $%
m-n $ \\ \hline\hline
$q_{L}$ & $=$ & $-n-p$ & $q_{E_{R}}$ & $=$ & $m+p$ & $q_{H_{u}}$ & $=$ & $m$
\\ \hline\hline
$q_{H_{d}}$ & $=$ & $-m+n$ &  &  &  &  &  &  \\ \hline\hline
\end{tabular}%
$%
\end{center}
\caption{Chiral fields and their discrete $g_{N}$ charges. }
\end{table}
\end{center}

The mixed anomaly $Z_{N}\times SU(3)^{2}$, $Z_{N}\times SU(2)^{2}$ and mixed
gravitational anomaly constraints applying the charge assignment (table 3)
read,

\begin{equation}
\begin{tabular}{cccc}
$SU(3)-SU(3)-Z_{N}$ & $:$ & $3n=0$ & mod $N,$ \\ 
$SU(2)-SU(2)-Z_{N}$ & $:$ & $2n+3p=0$ & mod $N,$ \\ 
$G-G-Z_{N}$ & $:$ & $-13n-3p+3m=0$ & mod $N+\eta \frac{N}{2},$%
\end{tabular}
\label{eq17}
\end{equation}%
where $\eta =0$ for $N$ being odd and $\eta =1$ for $N$ being even.

\section{Discrete gauge symmetries and Trinification model}

\subsection{$Z_{N}$ in intersecting D6-branes}

In type II orientifold constructions a stack of $N$ identical D6-branes
gives rise to a $U(N)$, that splits into $U(N)=SU(N)\times U(1)$ where the
abelian part is generically broken to $Z_{N}$ discrete gauge symmetries by
the presence of \ $N$ $B\wedge F_{\alpha }$ couplings, with $B$ being RR
2-form fields. Abelian discrete symmetries $Z_{N}$ are remnants of
continuous $U(1)$ symmetries that are broken by scalars with charge $N$
under the respective $U(1)$ acquiring vev's.

Consider a linear combination of the form $\sum\nolimits_{a}q_{a}U(1)_{a}$
of brane $a$, its $BF$ couplings are,

\begin{equation}
\begin{tabular}{ccc}
$\sum \left[ \frac{1}{2}\Sigma _{a}q_{a}N_{a}\left( \pi _{a}-\pi
_{a}^{\prime }\right) \right] \text{ }B_{2}$ & $\wedge $ & $F_{a},$%
\end{tabular}
\label{eq18}
\end{equation}%
where $\pi _{a}^{\prime }$ denotes the orientifold image cycle of $\pi _{a}.$
In \cite{14}, the authors give the criteria for the remaining massless
combination,

\begin{equation}
\begin{tabular}{ccc}
$\frac{1}{2}\Sigma _{a}q_{a}N_{a}\left( \pi _{a}-\pi _{a}^{\prime }\right) $
& $=$ & $0.$%
\end{tabular}
\label{eq19}
\end{equation}

In practice, it is convenient to work with a basis of three-cycles $\left\{
\alpha _{i}\right\} ,\left\{ \beta _{i}\right\} $ which are even and odd
under the orientifold action, with $i=0,...,h_{21}$, such that $\alpha _{k}$ 
$.\beta _{l}=\delta _{kl}.$ The wrapped cycles are expanded in terms of this
basis as,

\begin{equation}
\begin{tabular}{llllll}
$\pi _{a}$ & $=$ & $\sum_{i}\left( r_{a}^{i}\alpha _{i}+s_{i}^{i}\beta
_{i}\right) ,$ & $\pi _{a}^{\prime }$ & $=$ & $\sum_{i}\left(
r_{a}^{i}\alpha _{i}-s_{a}^{i}\beta _{i}\right) ,$%
\end{tabular}
\label{eq20}
\end{equation}%
where the coefficients $r_{a}^{i}$ and $s_{a}^{i}$ are integers and are
usually referred to as wrapping numbers, with,

\begin{equation}
\begin{tabular}{llllll}
$s_{a}^{0}$ & $=$ & $m_{a}^{1}m_{a}^{2}m_{a}^{3},$ & $s_{a}^{1}$ & $=$ & $%
m_{a}^{1}n_{a}^{2}n_{a}^{3},$%
\end{tabular}
\label{eq21}
\end{equation}

\begin{equation}
\begin{tabular}{llllll}
$\text{ }s_{a}^{2}$ & $=$ & $n_{a}^{1}m_{a}^{2}n_{a}^{3},$ & $s_{a}^{3}$ & $%
= $ & $n_{a}^{1}n_{a}^{2}m_{a}^{3}.$%
\end{tabular}
\label{eq22}
\end{equation}

The structure of the $BF$ coupling shows that a discrete $Z_{N}$ gauge
symmetry appears when the quantities $\sum_{a}q_{a}N_{a}s_{a}^{i}$ are
multiples of n, for all $i$. Under a $U(1)$ gauge transformation, the
constraint (\ref{eq19}) takes the form,

\begin{equation}
\begin{tabular}{lll}
$\sum_{a}\sum_{i}q_{a}N_{a}s_{a}^{i}\beta _{i}$ & $=$ & $0.$%
\end{tabular}
\label{eq23}
\end{equation}

We will assume that the three-cycles are orthogonal to each other (\ref{eq23}%
) reads,

\begin{equation}
\begin{tabular}{llll}
$\sum_{a}q_{a}N_{a}s_{a}^{i}$ & $=$ & $0$ & $\forall i.\text{ with }q_{a}\in 
%TCIMACRO{\U{211a} }%
%BeginExpansion
\mathbb{Q}
%EndExpansion
.$%
\end{tabular}
\label{eq24}
\end{equation}

This condition can be generalised for a discrete gauge symmetry $Z_{N}$
arising from a linear combination $Z_{N}=\sum_{i}k_{i}U(1)_{i}$ as,\ 

\begin{equation}
\begin{tabular}{lllll}
$\sum_{a}k_{a}N_{a}s_{a}^{i}$ & $=$ & $0$ & $\text{mod }N\text{ \ }\forall
i, $ & $\text{with }k_{a}\in 
%TCIMACRO{\U{2124} }%
%BeginExpansion
\mathbb{Z}
%EndExpansion
.$%
\end{tabular}
\label{eq25}
\end{equation}

\subsection{Trinification model with discrete symmetries}

We will now discuss the origin of discrete symmetries in trinification
model. To show how these symmetries appear we proceed further by imposing
the necessary condition (\ref{eq24}). The hypercharge remains massless as
long as,

\begin{equation}
\begin{tabular}{lll}
$m_{c1}n_{c2}n_{c3}+m_{l1}n_{l2}n_{l3}+m_{r1}n_{r2}n_{r3}$ & $=$ & $0,$ \\ 
&  &  \\ 
$n_{c1}n_{c2}m_{c3}+n_{l1}n_{l2}m_{l3}+n_{r1}n_{r2}m_{r3}$ & $=$ & $0,$ \\ 
&  &  \\ 
$m_{c1}m_{c2}m_{c3}+m_{l1}m_{l2}m_{l3}+m_{r1}m_{r2}m_{r3}$ & $=$ & $0,$ \\ 
&  &  \\ 
$n_{c1}m_{c2}n_{c3}+n_{l1}m_{l2}n_{l3}+n_{r1}m_{r2}n_{r3}$ & $=$ & $0,$%
\end{tabular}
\label{eq26}
\end{equation}%
where we have accounted for a factor of $N_{a}=3.$ In our model building, we
need to consider the constraints imposed on the winding numbers. We may use
the condition $I_{cc^{\ast }}=m_{c1}m_{c2}m_{c3}=0$ to eliminate the
existence of the symmetric and antisymmetric representations originated from
open strings with both end-points on $SU\left( 3\right) _{C},$ for example,
this condition can be satisfied by setting $m_{c2}=0,$ which imply,

\begin{equation}
\begin{tabular}{lll}
$m_{c1}n_{c2}n_{c3}+m_{l1}n_{l2}n_{l3}+m_{r1}n_{r2}n_{r3}$ & $=$ & $0,$ \\ 
&  &  \\ 
$n_{c1}n_{c2}m_{c3}+n_{l1}n_{l2}m_{l3}+n_{r1}n_{r2}m_{r3}$ & $=$ & $0,$ \\ 
&  &  \\ 
$m_{l1}m_{l2}m_{l3}+m_{r1}m_{r2}m_{r3}$ & $=$ & $0,$ \\ 
&  &  \\ 
$n_{l1}m_{l2}n_{l3}+n_{r1}m_{r2}n_{r3}$ & $=$ & $0.$%
\end{tabular}
\label{eq27}
\end{equation}

Using the above results and the conditions to obtain three fermion
generations, we solve for $m_{c1}$,

\begin{equation}
m_{c1}=-\frac{m_{l1}n_{c1}}{n_{l1}},  \label{eq28}
\end{equation}

which imply the relations,

\begin{equation}
\begin{tabular}{ccccc}
$I_{cl}$ & $=$ & $2m_{l2}n_{c2}m_{l1}n_{c1}\left(
m_{c3}n_{l3}-m_{l3}n_{c3}\right) $ & $=$ & $3,$%
\end{tabular}
\label{eq29}
\end{equation}%
and substitute to the conditions $I_{cr^{\ast }}=0$, this requires,

\begin{equation}
\begin{tabular}{lll}
$\left( m_{c1}n_{r1}+m_{r1}n_{c1}\right) \left(
m_{c3}n_{r3}+m_{r3}n_{c3}\right) $ & $=$ & $0,$%
\end{tabular}
\label{eq30}
\end{equation}%
and,

\begin{equation}
-\frac{n_{c1}}{n_{l1}}%
\begin{tabular}{lll}
$\left( m_{l1}n_{r1}-m_{r1}n_{l1}\right) \left(
m_{c3}n_{r3}+m_{r3}n_{c3}\right) $ & $=$ & $0,$%
\end{tabular}
\label{eq31}
\end{equation}%
we hence deduce,

\begin{equation}
\begin{tabular}{lll}
$m_{c3}n_{r3}$ & $=$ & $-m_{r3}n_{c3}.$%
\end{tabular}
\label{eq32}
\end{equation}%
Then with the conditions $I_{lr^{\ast }}=0$,

\begin{equation*}
\begin{tabular}{l}
$\left( m_{l1}n_{r1}+m_{r1}n_{l1}\right) \left(
m_{l2}n_{r2}+m_{r2}n_{l2}\right) \left( m_{l3}n_{r3}+m_{r3}n_{l3}\right) $%
\end{tabular}%
\end{equation*}

\begin{equation}
\begin{tabular}{ll}
$=$ & $0,$%
\end{tabular}
\label{eq33}
\end{equation}

and,

\begin{equation*}
-\frac{n_{c1}m_{r3}}{n_{l1}m_{c3}}%
\begin{tabular}{l}
$\left( m_{l2}n_{r2}+m_{r2}n_{l2}\right) \left(
m_{c1}n_{r1}-m_{r1}n_{c1}\right) \left( m_{c3}n_{l3}-m_{l3}n_{c3}\right) $%
\end{tabular}%
\end{equation*}

\begin{equation}
\begin{tabular}{l}
\begin{tabular}{ll}
$=$ & $0,$%
\end{tabular}%
\end{tabular}
\label{eq34}
\end{equation}

we therefore deduce,

\begin{equation}
\begin{tabular}{lll}
$m_{l2}n_{r2}$ & $=$ & $-m_{r2}n_{l2},$%
\end{tabular}
\label{eq35}
\end{equation}

\begin{equation}
\begin{tabular}{lll}
$m_{r1}n_{c1}m_{c3}n_{l3}$ & $=$ & $m_{l3}n_{c3}m_{c1}n_{r1},$%
\end{tabular}
\label{eq36}
\end{equation}

\begin{equation}
\frac{1}{m_{l2}}\left( m_{l2}m_{c1}n_{c2}n_{c3}+\frac{3n_{l2}}{%
2n_{c1}n_{c2}m_{c3}}\right) 
\begin{tabular}{ll}
$=$ & $0.$%
\end{tabular}
\label{eq37}
\end{equation}

Using all these, we get,

\begin{equation}
\frac{1}{n_{l1}m_{l2}}\left( \frac{9xy}{m_{l3}\left( 4xy+3\right) ^{2}}%
-yn_{c1}n_{c2}\right) 
\begin{tabular}{ll}
$=$ & $0,$%
\end{tabular}
\label{eq38}
\end{equation}%
and,

\begin{equation}
\begin{tabular}{ll}
$n_{l1}$ & $=$%
\end{tabular}%
\frac{-2xym_{l1}n_{r1}}{m_{r1}\left( 2xy+3\right) },  \label{eq39}
\end{equation}%
where $x=n_{c1}n_{c2}m_{l3}$ and $y=m_{l1}m_{l2}n_{c3}$.

Summarizing, we find,

\begin{equation}
\frac{4xy\left( 2xy+3\right) ^{2}m_{r1}}{\left( 4xy+3\right)
^{2}m_{l1}m_{l2}m_{l3}n_{r1}}%
\begin{tabular}{ll}
$=$ & $0,$%
\end{tabular}
\label{eq40}
\end{equation}

\begin{equation}
\frac{8x^{2}y\left( 2xy+3\right) m_{c3}}{\left( 4xy+3\right) ^{2}m_{l3}}%
\begin{tabular}{ll}
$=$ & $0,$%
\end{tabular}
\label{eq41}
\end{equation}%
we conclude that these conditions can be satisfied by setting $m_{r1}=0$ and 
$m_{l3}=0.$ For this particular solution, the computation of the conditions
for discrete gauge symmetries imposed on the winding numbers, requires the
results,

\begin{equation*}
n_{l1}=-\frac{m_{l1}n_{c1}}{m_{c1}},\text{ }n_{r3}=-\frac{m_{r3}n_{c3}}{%
m_{c3}},\text{ }n_{l3}=\frac{3}{2m_{l1}m_{l2}n_{c1}n_{c2}m_{c3}},
\end{equation*}

\begin{equation}
n_{l2}=-\frac{2}{3}m_{c1}n_{c1}n_{c2}^{2}n_{c3}m_{c3}m_{l2},\text{ }m_{r2}=-%
\frac{3}{2m_{c1}n_{c2}n_{c3}m_{r3}n_{r1}},  \label{eq42}
\end{equation}

\begin{equation*}
n_{r2}=-\frac{n_{c1}n_{c2}m_{c3}}{n_{r1}m_{r3}}.
\end{equation*}

Depending on the structure of the $B\wedge F$ couplings in the model, one
can identify the presence of a discrete $Z_{N}$ gauge symmetry, the relevant 
$BF$ couplings are,

\begin{equation}
\begin{tabular}{lll}
$F_{c}$ & $\wedge $ & $3\left(
m_{c1}n_{c2}n_{c3}B_{2}^{1}+n_{c1}n_{c2}m_{c3}B_{2}^{3}\right) ,$ \\ 
$F_{l}$ & $\wedge $ & $3\left(
m_{l1}n_{l2}n_{l3}B_{2}^{1}+n_{l1}m_{l2}n_{l3}B_{2}^{2}\right) ,$ \\ 
$F_{r}$ & $\wedge $ & $3\left(
n_{r1}m_{r2}n_{r3}B_{2}^{2}+n_{r1}n_{r2}m_{r3}B_{2}^{3}\right) .$%
\end{tabular}
\label{eq43}
\end{equation}

Using the results (\ref{eq42}), we end up with,

\begin{equation}
\begin{tabular}{lll}
$F_{c}$ & $\wedge $ & $3\left(
m_{c1}n_{c2}n_{c3}B_{2}^{1}+n_{c1}n_{c2}m_{c3}B_{2}^{3}\right) ,$ \\ 
$F_{l}$ & $\wedge $ & $3\left( -m_{c1}n_{c2}n_{c3}B_{2}^{1}-\frac{3}{%
2m_{c1}m_{c3}n_{c2}}B_{2}^{2}\right) ,$ \\ 
$F_{r}$ & $\wedge $ & $3\left( \frac{3}{2m_{c1}m_{c3}n_{c2}}%
B_{2}^{2}-n_{c1}n_{c2}m_{c3}B_{2}^{3}\right) .$%
\end{tabular}
\label{eq44}
\end{equation}

It is easy to see that this structure naturally contains the discrete gauge
symmetries:

the discrete symmetry $%
%TCIMACRO{\U{2124} }%
%BeginExpansion
\mathbb{Z}
%EndExpansion
_{3}$ appears whenever $m_{c1}m_{c3}n_{c2}=3$ and $n_{c1}n_{c2}m_{c3}=1.$%
Note that this $%
%TCIMACRO{\U{2124} }%
%BeginExpansion
\mathbb{Z}
%EndExpansion
_{3}$ symmetry is associated with the $U(1)_{r}$ brane.

the discrete symmetry $%
%TCIMACRO{\U{2124} }%
%BeginExpansion
\mathbb{Z}
%EndExpansion
_{N}$ appears whenever $m_{c1}n_{c2}n_{c3}=N$ and $n_{c1}n_{c2}m_{c3}$ is a
multiple of $N.$

\section{Conclusions}

In this work, we have interested in the discrete gauge symmetries in viable
models. We have considered a trinification model in the context of type IIA
orientifolds with intersecting D6-branes, and discussed the phenomenological
importance of $Z_{N}$ symmetries.

We have investigated the origin of $Z_{N}$ gauge symmetries from $B\wedge F$
couplings in trinification model $\mbox{U(3)}_{C}\times \mbox{U(3)}%
_{L}\times \mbox{U(3)}_{R}$. More precisely, we have considered the imposing
conditions on the winding numbers for the particular solution required for
the presence of a massless $U(1)$ in the low energy effective theory where
these discrete gauge symmetries appear.\ 

We hope the discussion of discrete gauge symmetries in D-brane model
buildings as done in this work is suffisant to show and motivate that they
still deserve more attention and more systematic understanding for future
works.

Acknowledgement: The authors would like to thank URAC09 CNSRT.

\end{document}